\documentclass[10pt,a4paper]{article}

\usepackage[utf8]{inputenc}
\usepackage[T1]{fontenc}
\usepackage{lmodern}
\usepackage[scaled=0.98]{zi4}
\usepackage[a4paper,top=20mm,bottom=23mm,left=22mm,right=22mm,headheight=15pt,footskip=11mm]{geometry}

\usepackage{authblk}
\usepackage[titletoc,title]{appendix}
\usepackage[dvipsnames]{xcolor}
\usepackage{color}
\usepackage{graphicx,epic,eepic,epsfig,latexsym,verbatim}
\usepackage[export]{adjustbox}
\usepackage{url}
\usepackage{xurl}
\usepackage{hyperref}
\hypersetup{colorlinks=true,citecolor=blue,linkcolor=blue,filecolor=blue,urlcolor=blue,breaklinks=true}
\usepackage{cite}

\usepackage{amsfonts,amsmath,amssymb,amsthm,mathtools,mathrsfs}
\usepackage{dsfont}
\usepackage{stmaryrd}
\usepackage{bm}
\usepackage[shortlabels]{enumitem}

\usepackage{framed}
\usepackage{subcaption}
\usepackage{caption}
\usepackage{float}
\usepackage{tikz}
\usetikzlibrary{arrows.meta,plotmarks,positioning}
\usepackage{pgfplots}
\pgfplotsset{compat=1.18}
\usepgfplotslibrary{patchplots}

\usepackage{microtype}
\usepackage{listings}
\usepackage[most]{tcolorbox}
\tcbuselibrary{skins,breakable,listings}
\usepackage{varwidth}
\usepackage{fancyhdr}
\usepackage{titlesec}
\usepackage{needspace}
\usepackage{relsize}
\usepackage{soul}
\soulregister\cite7
\soulregister\ref7
\soulregister\pageref7

\usepackage{colortbl}
\usepackage{pifont}
\usepackage{booktabs}
\usepackage{makecell}
\usepackage{diagbox}
\usepackage{multirow}
\usepackage{algorithm}
\usepackage{algorithmic}

\usepackage[marginal]{footmisc}
\usepackage{cleveref}

\newcounter{remark}

\newcounter{example}

\def\squareforqed{\hbox{\rlap{$\sqcap$}$\sqcup$}}
\def\qed{\ifmmode\squareforqed\else{\unskip\nobreak\hfil
      \penalty50\hskip1em\null\nobreak\hfil\squareforqed
      \parfillskip=0pt\finalhyphendemerits=0\endgraf}\fi}
\def\endenv{\ifmmode\;\else{\unskip\nobreak\hfil
      \penalty50\hskip1em\null\nobreak\hfil\;
      \parfillskip=0pt\finalhyphendemerits=0\endgraf}\fi}

\mathchardef\ordinarycolon\mathcode`\:
\mathcode`\:=\string"8000
\def\vcentcolon{\mathrel{\mathop\ordinarycolon}}
\begingroup \catcode`\:=\active
\lowercase{\endgroup
  \let :\vcentcolon
}

\definecolor{LightGray}{RGB}{220,220,220}
\definecolor{myred}{RGB}{236,17,0}
\definecolor{myblue}{RGB}{10,88,153}
\definecolor{myorange}{RGB}{236,137,0}
\definecolor{mygreen}{RGB}{26,152,81}
\definecolor{Gray}{gray}{0.92}
\definecolor{Gray2}{gray}{0.75}
\definecolor{maroon}{cmyk}{0,0.87,0.68,0.32}

\definecolor{QSTInk}{RGB}{31,40,54}
\definecolor{QSTNavy}{RGB}{24,55,91}
\definecolor{QSTMuted}{RGB}{102,116,135}
\definecolor{QSTPanelBlueDark}{RGB}{21,64,107}
\definecolor{QSTPanelBlueFrame}{RGB}{92,139,181}
\definecolor{QudeBlue}{RGB}{34,55,199}
\definecolor{QudeViolet}{RGB}{108,49,225}
\definecolor{QITLeanDecl}{RGB}{15,77,151}
\definecolor{QITLeanNamespace}{RGB}{16,108,137}
\definecolor{QITLeanProof}{RGB}{49,118,89}
\definecolor{QITLeanType}{RGB}{36,97,153}
\definecolor{QITLeanDomain}{RGB}{38,83,128}
\definecolor{QITLeanComment}{RGB}{94,111,124}
\definecolor{QITLeanString}{RGB}{26,118,116}
\definecolor{QITLeanWarn}{RGB}{145,69,60}

\DeclareRobustCommand{\BenchmarkHeader}{  \textnormal{Lean-QuantumAlg-Bench\enspace\textbar\enspace Lean-QIT-Bench}}
\newcommand{\QudeLeapLogo}[1]{  \includegraphics[#1]{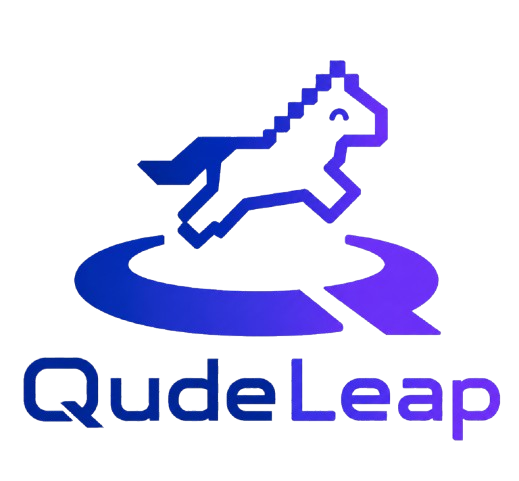}}

\titleformat{\section}
  {\Large\sffamily\bfseries\color{QSTPanelBlueDark}}
  {\thesection}{0.65em}{}
  [\vspace{0.15em}\color{QSTPanelBlueFrame!55}\titlerule]
\titleformat{\subsection}
  {\large\sffamily\bfseries\color{QSTNavy}}
  {\thesubsection}{0.65em}{}
\titleformat{\subsubsection}
  {\normalsize\sffamily\bfseries\color{QSTInk}}
  {\thesubsubsection}{0.65em}{}

\AddToHook{cmd/@floatboxreset/after}{\centering\setlength{\tabcolsep}{1em}}

\makeatletter
\renewcommand{\maketitle}{  \thispagestyle{empty}  \begingroup
  \setlength{\parindent}{0pt}  \noindent
  \begin{minipage}[t]{0.76\textwidth}
    \vspace{0pt}    {\raggedright\hyphenpenalty=10000\exhyphenpenalty=10000
      \sffamily\bfseries\color{QSTInk}\fontsize{23.5}{27.5}\selectfont
      \@title\par}  \end{minipage}  \hfill
  \begin{minipage}[t]{0.18\textwidth}
    \vspace{0pt}\raggedleft
    \QudeLeapLogo{width=2.65cm}  \end{minipage}
  \vspace{7mm}

  \noindent
  {\color{QudeBlue}\rule{0.22\textwidth}{1.8pt}}  {\color{QudeViolet}\rule{0.78\textwidth}{1.8pt}}\par
  \vspace{5mm}

  \begin{center}
    {\sffamily\normalsize\color{QSTInk}\@author\par}  \end{center}
  \vspace{3mm}
  \@thanks
  \endgroup
  \setcounter{footnote}{0}}
\makeatother

\lstdefinelanguage{Lean4}{
  sensitive=true,
  alsoletter={_},
  morekeywords=[1]{
    abbrev,class,def,deriving,example,inductive,instance,lemma,
    noncomputable,opaque,structure,theorem
  },
  morekeywords=[2]{
    end,export,namespace,open,private,protected,public,section,variable,
    variables,where
  },
  morekeywords=[3]{
    by,calc,change,do,else,exact,exists,for,forall,from,fun,have,if,
    inferInstance,infix,infixl,infixr,intro,intros,let,letI,macro,match,
    obtain,rcases,refine,rfl,rw,show,syntax,termination_by,then,with
  },
  morekeywords=[4]{
    Bool,False,Fin,Int,List,Nat,None,Option,Prop,Real,Set,Some,Sort,
    String,True,Type
  },
  morekeywords=[5]{axiom,sorry,admit},
  morekeywords=[6]{Benchmark,Dataset,Entry,Result},
  morecomment=[l]{--},
  morecomment=[s]{/-}{-/},
  morestring=[b]",
}

\lstdefinestyle{QudeLeanListing}{
  language=Lean4,
  basicstyle=\fontsize{9.8}{11.5}\selectfont\ttfamily\color{QSTInk},
  keywordstyle=[1]\color{QITLeanDecl}\bfseries,
  keywordstyle=[2]\color{QITLeanNamespace}\bfseries,
  keywordstyle=[3]\color{QITLeanProof},
  keywordstyle=[4]\color{QITLeanType}\bfseries,
  keywordstyle=[5]\color{QITLeanWarn}\bfseries,
  keywordstyle=[6]\color{QITLeanDomain}\bfseries,
  commentstyle=\color{QITLeanComment},
  stringstyle=\color{QITLeanString},
  numbers=none,
  xleftmargin=0pt,
  framexleftmargin=0pt,
  columns=fullflexible,
  keepspaces=true,
  showstringspaces=false,
  tabsize=2,
  breaklines=true,
  breakatwhitespace=false,
  breakindent=1.15em,
  breakautoindent=true,
  aboveskip=0pt,
  belowskip=0pt,
  literate=
    {_}{{\textunderscore\allowbreak}}1
    {∀}{{\ensuremath{\forall}}}1
    {∃}{{\ensuremath{\exists}}}1
    {→}{{\ensuremath{\to}}}1
    {←}{{\ensuremath{\leftarrow}}}1
    {↔}{{\ensuremath{\leftrightarrow}}}1
    {∧}{{\ensuremath{\wedge}}}1
    {∨}{{\ensuremath{\vee}}}1
    {¬}{{\ensuremath{\neg}}}1
    {≤}{{\ensuremath{\leq}}}1
    {≥}{{\ensuremath{\geq}}}1
    {≠}{{\ensuremath{\neq}}}1
    {⟨}{{\ensuremath{\langle}}}1
    {⟩}{{\ensuremath{\rangle}}}1
    {λ}{{\ensuremath{\lambda}}}1
    {α}{{\ensuremath{\alpha}}}1
    {β}{{\ensuremath{\beta}}}1
    {θ}{{\ensuremath{\theta}}}1
    {ℕ}{{\ensuremath{\mathbb{N}}}}1
    {ℤ}{{\ensuremath{\mathbb{Z}}}}1
    {ℚ}{{\ensuremath{\mathbb{Q}}}}1
    {ℝ}{{\ensuremath{\mathbb{R}}}}1
    {×}{{\ensuremath{\times}}}1
    {ˣ}{{\ensuremath{^{\times}}}}1
    {⁻¹}{{\ensuremath{^{-1}}}}2
    {χ}{{\ensuremath{\chi}}}1
    {–}{{--}}1
    {•}{{\ensuremath{\bullet}}}1
    {ℂ}{{\ensuremath{\mathbb{C}}}}1
    {ℓ}{{\ensuremath{\ell}}}1
    {↪}{{\ensuremath{\hookrightarrow}}}1
    {∈}{{\ensuremath{\in}}}1
    {∑}{{\ensuremath{\textstyle\sum}}}1
    {⊗}{{\ensuremath{\otimes}}}1
}

\newtcblisting[auto counter]{leanlisting}[3][]{  enhanced,
  breakable,
  listing only,
  listing engine=listings,
  listing options={style=QudeLeanListing},
  colback=gray!5,
  colframe=gray!55,
  boxrule=0.4pt,
  arc=0.6mm,
  left=4pt,
  right=4pt,
  top=4pt,
  bottom=4pt,
  title={Listing~\thetcbcounter: #2},
  fonttitle=\sffamily\small,
  coltitle=black,
  attach boxed title to top left={yshift=-2mm,xshift=4mm},
  varwidth boxed title=\dimexpr\linewidth-8mm\relax,
  boxed title style={colback=gray!18,sharp corners,boxrule=0pt,boxsep=2pt},
  label={#3},
  before skip=0.6\baselineskip,
  after skip=0.65\baselineskip,
  #1
}

\newtcbinputlisting[use counter from=leanlisting]{\leaninputlisting}[3]{  enhanced,
  breakable,
  listing only,
  listing engine=listings,
  listing file={#3},
  listing options={style=QudeLeanListing},
  colback=gray!5,
  colframe=gray!55,
  boxrule=0.4pt,
  arc=0.6mm,
  left=4pt,
  right=4pt,
  top=4pt,
  bottom=4pt,
  title={Listing~\thetcbcounter: #1},
  fonttitle=\sffamily\small,
  coltitle=black,
  attach boxed title to top left={yshift=-2mm,xshift=4mm},
  varwidth boxed title=\dimexpr\linewidth-8mm\relax,
  boxed title style={colback=gray!18,sharp corners,boxrule=0pt,boxsep=2pt},
  label={#2},
  before skip=0.6\baselineskip,
  after skip=0.65\baselineskip
}

\DeclareRobustCommand{\LeanQIT}{\textnormal{\textsc{Lean-QIT}}}

\DeclareUnicodeCharacter{00B7}{\ensuremath{\cdot}}
\DeclareUnicodeCharacter{03A6}{\ensuremath{\Phi}}
\DeclareUnicodeCharacter{03A8}{\ensuremath{\Psi}}
\DeclareUnicodeCharacter{03B1}{\ensuremath{\alpha}}
\DeclareUnicodeCharacter{03B2}{\ensuremath{\beta}}
\DeclareUnicodeCharacter{03B3}{\ensuremath{\gamma}}
\DeclareUnicodeCharacter{03B4}{\ensuremath{\delta}}
\DeclareUnicodeCharacter{03B5}{\ensuremath{\varepsilon}}
\DeclareUnicodeCharacter{03B7}{\ensuremath{\eta}}
\DeclareUnicodeCharacter{03B8}{\ensuremath{\theta}}
\DeclareUnicodeCharacter{03B9}{\ensuremath{\iota}}
\DeclareUnicodeCharacter{03BA}{\ensuremath{\kappa}}
\DeclareUnicodeCharacter{03BB}{\ensuremath{\lambda}}
\DeclareUnicodeCharacter{03BC}{\ensuremath{\mu}}
\DeclareUnicodeCharacter{03C0}{\ensuremath{\pi}}
\DeclareUnicodeCharacter{03C1}{\ensuremath{\rho}}
\DeclareUnicodeCharacter{03C3}{\ensuremath{\sigma}}
\DeclareUnicodeCharacter{03C4}{\ensuremath{\tau}}
\DeclareUnicodeCharacter{03C6}{\ensuremath{\phi}}
\DeclareUnicodeCharacter{03C8}{\ensuremath{\psi}}
\DeclareUnicodeCharacter{03C9}{\ensuremath{\omega}}
\DeclareUnicodeCharacter{2115}{\ensuremath{\mathbb{N}}}
\DeclareUnicodeCharacter{211D}{\ensuremath{\mathbb{R}}}
\DeclareUnicodeCharacter{2192}{\ensuremath{\to}}
\DeclareUnicodeCharacter{2194}{\ensuremath{\leftrightarrow}}
\DeclareUnicodeCharacter{21A6}{\ensuremath{\mapsto}}
\DeclareUnicodeCharacter{21AA}{\ensuremath{\hookrightarrow}}
\DeclareUnicodeCharacter{2200}{\ensuremath{\forall}}
\DeclareUnicodeCharacter{2203}{\ensuremath{\exists}}
\DeclareUnicodeCharacter{2208}{\ensuremath{\in}}
\DeclareUnicodeCharacter{2209}{\ensuremath{\notin}}
\DeclareUnicodeCharacter{2211}{\ensuremath{\sum}}
\DeclareUnicodeCharacter{221E}{\ensuremath{\infty}}
\DeclareUnicodeCharacter{2227}{\ensuremath{\wedge}}
\DeclareUnicodeCharacter{2228}{\ensuremath{\vee}}
\DeclareUnicodeCharacter{2282}{\ensuremath{\subset}}
\DeclareUnicodeCharacter{2286}{\ensuremath{\subseteq}}
\DeclareUnicodeCharacter{2293}{\ensuremath{\sqcap}}
\DeclareUnicodeCharacter{2294}{\ensuremath{\sqcup}}
\DeclareUnicodeCharacter{2297}{\ensuremath{\otimes}}
\DeclareUnicodeCharacter{22A4}{\ensuremath{\top}}
\DeclareUnicodeCharacter{22A5}{\ensuremath{\bot}}
\DeclareUnicodeCharacter{2264}{\ensuremath{\le}}
\DeclareUnicodeCharacter{2265}{\ensuremath{\ge}}
\DeclareUnicodeCharacter{2260}{\ensuremath{\ne}}
\DeclareUnicodeCharacter{00D7}{\ensuremath{\times}}
\DeclareUnicodeCharacter{27E8}{\ensuremath{\langle}}
\DeclareUnicodeCharacter{27E9}{\ensuremath{\rangle}}

\newcommand{\bra}[1]{\langle#1|}
\newcommand{\ket}[1]{|#1\rangle}

\newcommand{\proj}[1]{| #1\rangle\!\langle #1 |}

\newcommand{\tr}{\operatorname{Tr}}

\newcommand{\cH}{{\cal H}}

\newcommand{\cL}{{\cal L}}

\newcommand{\cS}{{\cal S}}

\makeatletter
\newcommand*\rel@kern[1]{\kern#1\dimexpr\macc@kerna}
\newcommand*\widebar[1]{  \begingroup
  \def\mathaccent##1##2{    \rel@kern{0.8}    \overline{\rel@kern{-0.8}\macc@nucleus\rel@kern{0.2}}    \rel@kern{-0.2}  }  \macc@depth\@ne
  \let\math@bgroup\@empty \let\math@egroup\macc@set@skewchar
  \mathsurround\z@ \frozen@everymath{\mathgroup\macc@group\relax}  \macc@set@skewchar\relax
  \let\mathaccentV\macc@nested@a
  \macc@nested@a\relax111{#1}  \endgroup
}
\makeatother

\newcommand{\cptp}{\text{\rm CPTP}}

\NewDocumentCommand{\ddx}{o m}{  \IfNoValueTF{#1}
  {\frac{\operatorname{d}}{\operatorname{d}\!#2}}
  {\frac{\operatorname{d}^{\,#1}}{\operatorname{d}\!#2^{\,#1}}}}

\title{Benchmarking Agents for Proving Theorems in Quantum Algorithms and Quantum Information}

\author[1,2]{Lei Zhang\thanks{Equal contribution. leizhang116.4@gmail.com}}
\author[1,2]{Yusheng Zhao\thanks{Equal contribution. yushengzhao2020@outlook.com}}
\author[2]{Yimeng Cao}
\author[3]{Ranyiliu Chen}
\author[1,2]{Mingrui Jing\thanks{mingruij0031@gmail.com}}
\author[2]{Jizhe Lai}
\author[2]{Ziao Tang}
\author[2]{Jingu Xie}
\author[1,2]{Hongshun Yao}
\author[1,4]{Xuanqiang Zhao\thanks{zhaoxuanqiang7@gmail.com}}
\author[2]{Guocheng Zhen}
\author[1]{Chengkai Zhu\thanks{zhuchengkai7@gmail.com}}
\author[2]{Xin Wang\thanks{felixxinwang@hkust-gz.edu.cn}}
\affil[1]{\small QudeLeap Research, Shanghai 200030, China}
\affil[2]{\small The Hong Kong University of Science and Technology (Guangzhou), Guangdong 511453, China}
\affil[3]{\small Quantum Science Center of Guangdong-Hong Kong-Macao Greater Bay Area, Shenzhen 518045, China}
\affil[4]{\small The University of Hong Kong, Pokfulam Road, Hong Kong}
\date{}

\begin{document}
\maketitle

\vspace{-5mm}
\begin{center}
  \small\sffamily\color{QSTMuted}  \raisebox{-0.5ex}{\QudeLeapLogo{height=3.2mm}}\hspace{4pt}  \href{https://github.com/QudeLeap/Lean-QuantumAlg-Bench}{Lean-QuantumAlg-Bench}  \quad\textbar\quad
  \href{https://github.com/QuAIR/Lean-QIT-Bench}{Lean-QIT-Bench}\end{center}

\begin{abstract}
Formal verification is becoming increasingly practical for quantum computing, yet the ability of AI agents to construct machine-checkable proofs in this domain remains unmeasured. We introduce \textbf{Lean-QuantumAlg-Bench} and \textbf{Lean-QIT-Bench}, two Lean~4 benchmarks containing 36 and 40 theorem-completion tasks for quantum algorithms and quantum information theory, respectively. Every task compiles in a fixed environment and is evaluated by deterministic proof checking and targeted semantic review, with difficulty weights assigned before model execution. We evaluate four models---GPT-5.5, Kimi~K3, DeepSeek~V4-Pro, and MiniMax~M3---within a common theorem-proving framework under two settings: a task-only baseline and library-augmented deduction (LAD), which additionally provides access to a verified domain library. The highest difficulty-weighted scores are 60.4 out of 100 on the quantum-algorithm benchmark and 59.6 out of 100 on the quantum-information benchmark. LAD improves both score and completion rate in all eight model--benchmark comparisons, with gains of up to 15.9 points, providing evidence that verified libraries can strengthen domain-specific proof agents. The results reveal recurring weaknesses of agentic proving in areas such as quantum simulation, quantum learning, quantum information measures, and entanglement theory. Monetary and wall-clock costs per score point also vary substantially across models, highlighting important capability--efficiency trade-offs. We expect these benchmarks to establish a reproducible baseline for developing more capable and reliable proof agents, and to pave the way toward self-evolving AI scientists for advancing quantum information science.
\end{abstract}

\section{Introduction}
\label{sec:intro}

Quantum computing combines mathematically precise claims with long chains of
domain-specific reasoning.  Shor's factoring algorithm~\cite{shor1994algorithms}, for
example, reduces factoring to finding the multiplicative order of an element
modulo~$N$; a quantum subroutine recovers this order from the period of modular
exponentiation.  In quantum information theory, the data-processing
inequality~\cite{lindblad1975completely,wilde2017quantum} states that quantum relative entropy
cannot increase under a quantum channel.  Both results
connect abstract hypotheses, typed quantum objects, and operational
conclusions.  A machine-checkable development can make every link in such a
chain available for verification, including assumptions that a conventional
derivation may leave implicit.

Formalizing these arguments is demanding for reasons that are intrinsic to the
subject.  Quantum states and operators carry finite-dimensional and often
type-level structure; circuit calculations must connect syntactic
transformations with linear-algebraic semantics; and information-theoretic
inequalities depend on domains, support conditions, and positivity hypotheses.
Textbook notation also suppresses coercions, basis choices, tensor-factor
ordering, and index conventions.  In Lean, each of these choices becomes part
of the theorem interface or its imported environment.  A useful domain
benchmark should therefore test more than isolated algebra: it should reveal
whether a prover can navigate the interfaces through which the mathematics is
actually represented.

Formal verification has already reached substantial quantum results, including
certified quantum computation in Isabelle/HOL~\cite{bordg2021certified}, an end-to-end
implementation of Shor factorization~\cite{peng2023formally}, and a Lean formalization
of the generalized quantum Stein lemma~\cite{meiburg2025formalization}.
Recent preprints and libraries further develop Lean interfaces for
quantum information~\cite{leanquantum,leanqit}, algorithms~\cite{zhang2026building, jing2026agentic}, error correction~\cite{ehatamm2026endtoend}, and optimization problems~\cite{kol2026machineverified}.
These developments show that proof assistants can support deep quantum mathematics.  
Meanwhile, formal theorem proving in Lean is increasingly organized as an
agentic loop combining retrieval, decomposition, iterative repair, and verifier
feedback. Representative systems include COPRA~\cite{thakur2024copra}, AlphaProof~\cite{hubert2026alphaproof}, DeepSeek-Prover~\cite{xin2024deepseekprover15,ren2025deepseekproverv2},
Numina-Lean-Agent~\cite{liu2026numinaleanagent}, A Minimal Agent~\cite{requena2026minimalagent}, and Goedel-Architect~\cite{chung2026goedelarchitect}. For research-level formalization, Archon uses LeanSearch for task decomposition,
iterative refinement, and proof synthesis~\cite{ju2026automatedconjecture}, with LeanSearch v2 extending this line toward global premise retrieval~\cite{gao2026leansearchv2}.

Formal theorem-proving benchmarks have advanced in parallel.  Competition
benchmarks such as miniF2F~\cite{minif2f} and
PutnamBench~\cite{putnambench} test whole-proof generation on
standardized problems.  LeanDojo~\cite{leandojo} makes library
retrieval an explicit part of proving, while
miniCTX~\cite{minictx} evaluates prover use of long project
contexts.  Ref.~\cite{ospanov2025minif2flean} documents
informal--formal fidelity failures, while
Ref.~\cite{ammanamanchi2026faults} shows that statement defects and
evaluation-harness failures can distort benchmark scores.
Together, these efforts provide strong methods for evaluating proof search,
context use, and benchmark quality.  We build on these methods to coordinate
Lean evaluation across quantum algorithms and quantum information.

A coordinated benchmark for these domains should satisfy several requirements
simultaneously.  It
should cover domain objects and proof patterns beyond the scope of generic
mathematical suites and present the same theorem-completion task format under
controlled evaluation conditions.  Its construction claims must also match the
evidence that exists: Lean compilation and deterministic checks apply to every
task, whereas manual comparison with the mathematical source is a targeted
safeguard for selected or high-risk tasks.  This distinction limits score
interpretation to agent performance under the stated checks.

In this work, we introduce \textbf{Lean-QuantumAlg-Bench} (QAlg-Bench) and \textbf{Lean-QIT-Bench} (QIT-Bench), two coordinated Lean~4 benchmark suites containing 36 and 40 theorem-completion tasks for quantum algorithms and quantum information theory, respectively. Their 76 tasks are organized into six fields: three for quantum algorithms and three for quantum information. Both suites use a common theorem-completion contract and the same deterministic proof checks, while retaining domain-specific definitions and statements. We evaluate four models within a common theorem-proving framework under two settings: a task-only baseline and library-augmented deduction (LAD), which additionally provides access to a verified domain library for consultation alongside the shared theorem-completion task format.

The evaluated tasks provide no hints. This design allows us to examine how difficulty-weighted verification scores vary across models, fields, and library-access conditions, as well as the monetary and full-suite wall-clock costs per score point associated with these outcomes. These controlled single-run measurements establish a reproducible baseline for developing more capable and reliable proof agents, and pave the way toward self-evolving AI scientists for advancing quantum information science.
\section{Background and related work}
\label{sec:background-related-work}

Lean~\cite{moura2015lean,moura2021lean} is an interactive theorem prover and a
functional programming language based on dependent type theory.  A declaration is
accepted only after elaboration produces a proof term that the trusted kernel
checks against its type.  Elaborators, tactics, and automation may search for
that term, but they do not enlarge the kernel's notion of correctness.  For an
agent, the operative task is therefore defined by a theorem statement and its
imported environment: proposed terms or tactics transform an evolving proof
state, and success means that the resulting declaration compiles in the fixed
Lean environment.  The Lean Mathematical Library (MathLib)~\cite{mathlib}
supplies much of the reusable mathematical interface on which such tasks
depend.  Proof generation, premise selection, retrieval, and repository
interaction address complementary parts of the task.

Quantum formalization adds a distinctive layer of typed structure to this
interface.  States and effects are represented through matrices or operators;
measurements and channels impose positivity and normalization conditions;
circuit syntax must be related to denotational semantics; and entropy,
distinguishability, and coding statements introduce additional analytic
hypotheses.  Algorithmic developments further require finite groups,
polynomials, phase transformations, or simulation primitives.  Existing
formal work demonstrates several parts of this range, from certified circuit
semantics and Shor factorization to quantum-information
theorems~\cite{bordg2021certified,peng2023formally,meiburg2025formalization}.  Recent Lean developments
include Lean-Quantum, an infrastructure aimed at AI-assisted formalization of
quantum information~\cite{leanquantum}; the reusable
finite-dimensional quantum-information interfaces and coding-theorem endpoints
of Lean-QIT~\cite{leanqit}; and a formalization of the Shor algorithm
family spanning order finding and reversible modular and elliptic-curve
arithmetic~\cite{zhang2026building}.
The resulting library interfaces determine the actual proof obligations: two
informal formulas that look identical may elaborate differently because their
types, tensor conventions, or imported definitions differ.

Several early formal-reasoning benchmarks standardized task statements and
criteria for proof acceptance.
miniF2F~\cite{minif2f} translated olympiad-style problems across
proof assistants, while PutnamBench~\cite{putnambench} extended
competition evaluation to university-level Putnam problems and multiple
formal systems.  MLFMF~\cite{mlfmf} instead derived Lean and
Agda datasets for formalization recommendation from existing libraries.
Lean Workbook~\cite{leanworkbook} addressed data scarcity through a
large generated-and-filtered collection of informal/formal pairs.  These
resources differ in whether the unit
is a hand-formalized problem, a library declaration, or a generated pair, but
each uses fixed formal targets and machine-checkable acceptance.

Later work highlights two limitations of statement-only evaluation that are
relevant here: the role of context and the validity of the target itself.
LeanDojo~\cite{leandojo} combines programmatic repository
interaction, fine-grained premise annotations, and retrieval-augmented
proving.  miniCTX~\cite{minictx} evaluates theorem proving over long
project contexts with in-file dependencies.
RLMEval~\cite{rlmeval} selects
research-level theorems from Lean blueprint projects, while
Ineq-Comp~\cite{ineqcomp} probes robustness under transformed and
composed inequalities.
Ref.~\cite{ospanov2025minif2flean} reports that formally accepted statements can
still diverge from the intended informal problems.  In neighboring domain-specific
evaluation, QCircuitBench~\cite{qcircuitbench} tests quantum-algorithm
code design with automatic validation.  Ref.~\cite{cambench} extends
Lean evaluation to computational and applied mathematics with explicit
attention to dependencies and semantic review.
Ref.~\cite{ammanamanchi2026faults} documents how statement defects and
evaluation harnesses can distort formal-proving results.

These developments motivate the combination adopted here.  Competition and
research benchmarks supply standardized formal targets; repository-level work
shows why imported context and library access must be controlled; and
quality audits show why kernel acceptance cannot by itself certify the intended
meaning of a benchmark statement.  Our design coordinates these strands across
quantum algorithms and quantum information, partitions their mathematical
scope into interpretable fields, and evaluates a fixed task format under
baseline and LAD access.  QAlg-Bench and QIT-Bench use shared deterministic checks and
targeted semantic safeguards.  The next section explains their construction;
Section~\ref{sec:evaluation} reports their descriptive evaluation, and
Sections~\ref{sec:lean-quantumalg-bench} and~\ref{sec:lean-qit-bench} then
describe their field structure and representative tasks.
\section{Benchmark construction and validation}
\label{sec:benchmark-construction-validation}

\subsection{Task design and benchmark construction}
\label{sec:task-design-construction}

\begin{figure}[t]
  \centering
  \includegraphics[width=0.8\linewidth]{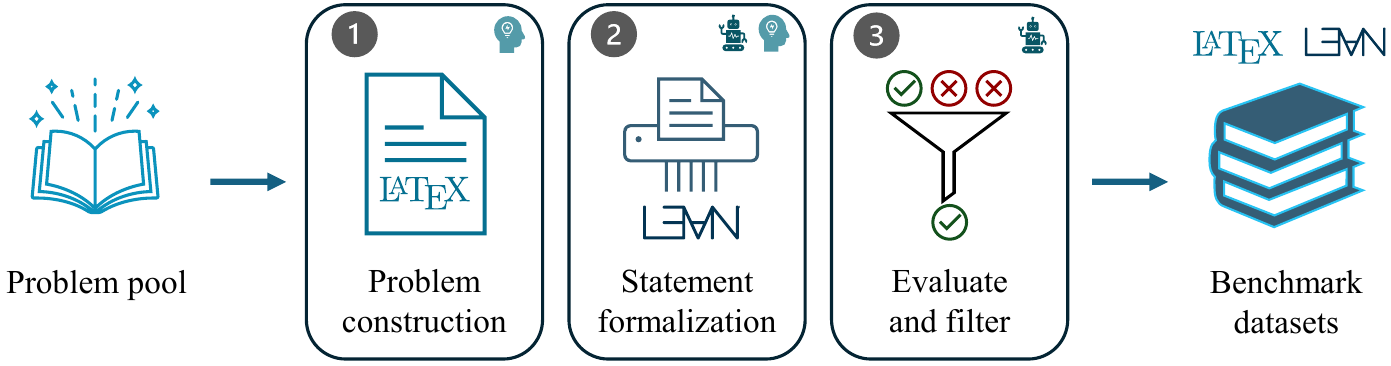}
  \caption{Benchmark construction workflow. Candidate problems are selected
  and written, translated into Lean, and checked before they enter the
  QAlg-Bench and QIT-Bench suites.}
  \label{fig:benchmark-overview}
\end{figure}

In this section, we construct QAlg-Bench and QIT-Bench following the workflow shown in Figure~\ref{fig:benchmark-overview}. Researchers first convert source problems into theorem-sized tasks. Agents assist in translating these statements into Lean, while researchers retain responsibility for the mathematical choices. Automated checks then compile and screen each task before it is included in one of the two benchmark suites.

The problem pool covers topics treated in established literature on quantum
information theory and quantum algorithms.  The quantum-information tasks
include material treated in standard and specialist texts of the
field~\cite{nielsen2010quantum,wilde2017quantum,tomamichel2016quantum}, together
with foundational results in quantum coding, channel capacity, entropy, and
entanglement~\cite{hayashi2003general,lloyd1997capacity,schumacher1995quantum,holevo1998capacity,bennett1996purification,alicki2004continuity,winter2016tight}.
The quantum-algorithm tasks cover canonical algorithms and simulation
techniques represented by Refs.~\cite{shor1994algorithms,bernstein1997quantum,childs2012hamiltonian,gilyen2019quantum}.

Candidates are selected for scientific relevance, theorem-sized scope, and a
viable Lean representation.  They originate from established quantum results,
library development objectives, or source statements whose assumptions can be
made explicit.  Duplicates, purely auxiliary declarations, and claims that
cannot be expressed within the fixed task format are excluded.  The admitted
tasks are then assigned to one of six fields.  QAlg-Bench uses state and operator
methods (SOM), circuit and algebraic algorithms (CAA), and simulation, signal
processing, and learning (SSL).  QIT-Bench uses quantum channels and representations
(QCR), operator and state geometry, symmetry, and distinguishability (GSD), and
quantum information measures and entanglement (IME).  Membership in a field is
fixed by the task's scope, and each field's difficulty denominator is the sum
of its task difficulties.  The complete inventories appear in
Appendix~\ref{app:full-benchmark-inventories}.

A Lean task contains a theorem statement and the supporting definitions needed
to state it.  Researchers determine the mathematical scope, assumptions, and
conclusion, and agents assist with the Lean translation.  The definitions
introduce benchmark-local objects that are not supplied by imported libraries,
and the statement contains the theorem signature whose body the agent must
complete.  The evaluated tasks provide no hints.

Every admitted task passes the same automated checks.  Its statement and
supporting definitions are assembled in the fixed Lean environment and
compiled by Lean before inclusion in the benchmark.
This checking stage applies to every task and denotes automated compilation and
screening; it does not imply that compilation guarantees mathematical fidelity
or that every statement receives the targeted manual comparison described
below.
Submission-side acceptance applies the same discipline: the independent
final verification recompiles the submitted theorem in the fixed
environment and accepts it only when the proof introduces no new axioms,
contains no \texttt{sorry} placeholder, and leaves all files outside the
theorem body unchanged.

The same benchmark task is used for both evaluation conditions.  The baseline
condition supplies the statement and supporting definitions.  LAD supplies the
same task together with a verified collection of library references available
only for consultation.  Each result records a fixed suite version,
model setting, access condition, and execution setting, which support
descriptive aggregation at the run level.

These construction steps establish the common mechanical checks.  The next
subsection addresses the separate question of whether selected formal
statements faithfully represent their motivating mathematics.
\subsection{Quality assurance and semantic validation}
\label{sec:quality-assurance-semantic-validation}

Lean's kernel guarantees that an accepted proof has the stated theorem as its type, relative to the imported environment~\cite{moura2015lean,moura2021lean}. It does not, however, guarantee that the statement faithfully captures the scientific claim that motivated the task. A benchmark can therefore be mechanically valid while remaining semantically misleading: an omitted hypothesis, an incorrect type or function encoding, or a weakened conclusion can cause the formal target to diverge from the intended mathematics~\cite{ospanov2025minif2flean}. We therefore treat Lean compilation and the automated checks applied to every task as universal mechanical guarantees, while regarding semantic fidelity as a separate validity question.

For selected statements whose translation carries elevated semantic risk, we perform a manual comparison between the formal signature, its supporting definitions, and the available mathematical specification. This comparison examines the relevant objects and quantifiers, explicit and implicit assumptions, dimensional constraints, index and tensor ordering, and the direction and strength of the conclusion. Any discrepancies identified in this process are resolved before the task is included in the benchmark.

This targeted comparison supplements the automated checks applied to every task. Because it is performed only on selected high-risk statements, it is not intended to define a benchmark-wide rate of semantic validity. With the benchmark construction, automated checks, and targeted semantic safeguards in place, the next section evaluates the recorded model runs.

\section{Evaluation}
\label{sec:evaluation}

The evaluation addresses three complementary questions: how objective proof
completion varies across models and access conditions, what resource cost
accompanies the observed scores, and whether suite-level aggregates conceal
field-specific strengths or weaknesses.

\begin{figure}[t]
  \centering
  \includegraphics[width=1.02\linewidth]{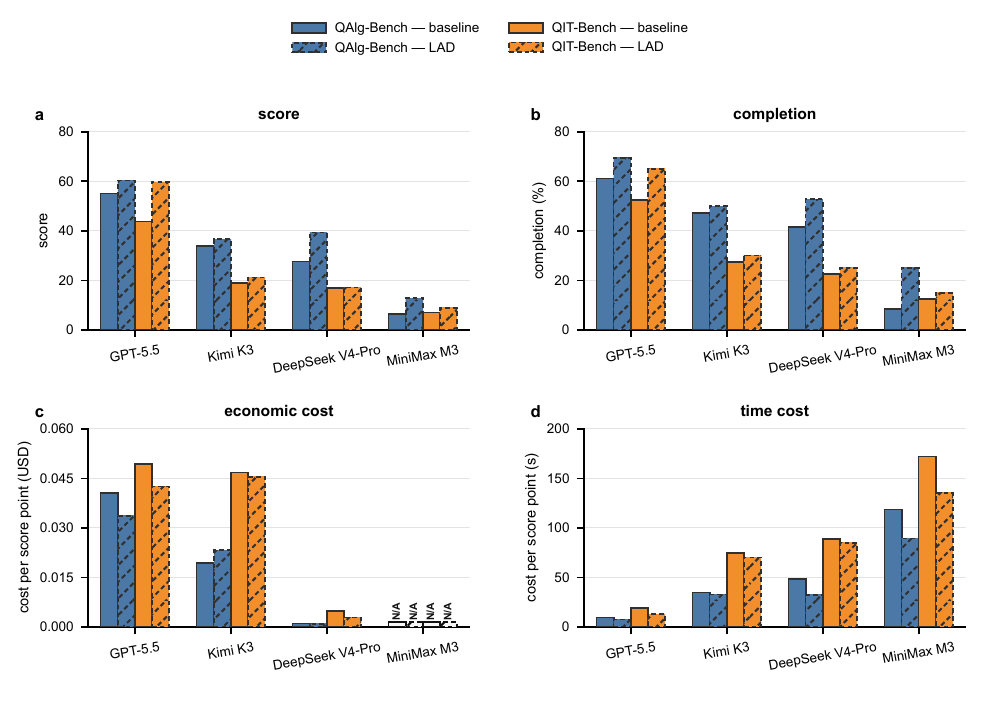}
  \caption{Verified performance and per-score costs across QAlg-Bench and QIT-Bench.
  Panels (a,b) show the difficulty-weighted score and completion, respectively.
  Panels (c,d) show economic cost in USD per score point and time cost in
  seconds per score point, respectively; lower values indicate lower cost.
  Blue and orange identify QAlg-Bench and QIT-Bench.  Solid, unhatched bars indicate the
  baseline condition, while dashed, diagonally hatched bars indicate LAD.
  Black N/A labels mark unavailable economic-cost values.}
  \label{fig:evaluation-summary}
\end{figure}

We evaluate one recorded run for each suite--model--condition combination.
Every run records its suite version and problem set, model and agent setting,
access condition, budget, retry policy, and execution environment.  These
settings are fixed within a run, although the historical baseline and LAD pairs
are not fully matched across all execution details.  QAlg-Bench contributes 36
tasks and QIT-Bench contributes 40, giving 16 combinations.
The runs also differ in agent tool: GPT-5.5 uses a different agent tool
from the other three models (Table~\ref{tab:model-settings}), so each
combination measures a model and agent tool jointly, and cross-model orderings
compare these joint settings.
The exact model and agent settings are listed in
Appendix~\ref{app:full-benchmark-inventories}.

For task $i$, let $d_i\in\{1,\ldots,10\}$ be the difficulty fixed before model
execution, and let $v_i\in\{0,1\}$ be the recorded acceptance indicator for the
final submitted theorem under the common Lean acceptance checks.  We define the score as
\begin{equation}
  \mathrm{score}=100\frac{\sum_i d_i v_i}{\sum_i d_i}.
  \label{eq:objective-score}
\end{equation}
Completion is the unweighted fraction of tasks with $v_i=1$.  The score therefore
weights verified tasks by their pre-exam difficulty, whereas completion treats
every task equally.  A model-call timeout does not itself set $v_i=0$: after
the invocation ends, an independent Lean verification checks the final
submission.  Four tasks in the snapshot reached the timeout and nevertheless
passed this final check.  Neither metric assesses intermediate proof progress.

Economic cost is the mean API list-price-equivalent USD among tasks with
reconstructable prices, divided by the difficulty-weighted score for those
tasks.  Time cost is the mean model invocation time per task over the complete
suite
divided by the overall score.
Model invocation time is the elapsed duration of the complete agent invocation,
including Lean checks initiated by the agent; it excludes the independent
final verification performed after the invocation ends.  The general pricing
schedule was fixed on 2026-07-20, while the Kimi K3 prices were fixed on
2026-07-22.  The units are USD and seconds per score point, and lower values
indicate lower cost.  Both ratios require a positive score denominator;
unavailable ratios are reported as N/A, with zero reserved for observed zeros.

\subsection{Overall performance}
\label{sec:evaluation-overall}

Figure~\ref{fig:evaluation-summary}(a,b) compares the 16 recorded results.
GPT-5.5 has the highest observed score in every suite--condition combination.
In all eight historical within-model comparisons, the LAD result has higher
score and completion than the baseline result.  The magnitude varies
substantially: DeepSeek V4-Pro, for example, shows a much larger separation
between conditions on QAlg-Bench than on QIT-Bench.  These single-run differences describe the
recorded runs; the unmatched execution details prevent their interpretation
as estimates of an access-condition effect.

Figure~\ref{fig:evaluation-summary}(c,d) shows that the cost ordering differs
from the performance ordering.  DeepSeek V4-Pro has the lowest reported
economic cost, while GPT-5.5 has the lowest observed time cost.  These
quantities are costs per score point.  Total expenditure, total elapsed time,
and intrinsic model speed are separate quantities.

\begin{table}[t]
  \caption{Relative change from baseline to LAD by suite and model.  Every
  metric column uses $(L-B)/B$ computed from unrounded values.  Positive score
  and completion values indicate higher verified performance; positive
  economic- and time-cost values indicate higher cost per score point, and
  negative values indicate lower cost per score point.  Economic cost divides
  mean API list-price-equivalent USD among tasks with reconstructable prices
  by the corresponding score.  Time cost divides mean model invocation seconds
  per task
  over the full suite by the overall score.  A black em dash denotes an
  unavailable observed metric.  All eight observed score changes are positive,
  but their magnitudes and the cost changes vary by model and suite.}
  \label{tab:evaluation-paired}
\begin{tabular}{@{}llrrrr@{}}
\toprule
Suite & Model & Score & Completion & Economic cost & Time cost \\
\midrule
QAlg-Bench & GPT-5.5 & +9.6\% & +13.6\% & -17.0\% & -19.5\% \\
 & Kimi K3 & +7.8\% & +5.9\% & +19.9\% & -6.1\% \\
 & DeepSeek V4-Pro & +42.5\% & +26.7\% & -11.1\% & -33.1\% \\
 & MiniMax M3 & +100.0\% & +200.0\% & \textemdash{} & -24.8\% \\
\midrule
QIT-Bench & GPT-5.5 & +36.4\% & +23.8\% & -13.7\% & -32.9\% \\
 & Kimi K3 & +11.3\% & +9.1\% & -2.8\% & -6.0\% \\
 & DeepSeek V4-Pro & +1.8\% & +11.1\% & -40.4\% & -4.0\% \\
 & MiniMax M3 & +26.1\% & +20.0\% & \textemdash{} & -21.4\% \\
\bottomrule
\end{tabular}
\end{table}

Table~\ref{tab:evaluation-paired} expresses each available LAD--baseline
difference relative to its baseline value as $(L-B)/B$, using unrounded
condition-level values.  Positive score or completion entries denote a higher
LAD value.  For economic and time cost, positive entries denote higher cost per
score point and negative entries denote lower cost per score point.  The
economic-cost entries compare the corresponding condition-level ratios.  A
black em dash marks an unavailable observed comparison.

Relative changes should be read together with their absolute baseline values.
MiniMax M3 doubles its QAlg-Bench score from 6.4 to 12.8 from a low baseline,
while the larger absolute increases for GPT-5.5 on QIT-Bench and DeepSeek
V4-Pro on QAlg-Bench
are 15.9 and 11.7 score points.  These ratios describe the recorded historical
runs; matched reruns are required to isolate the access-condition difference.
The absolute values are reported in
Appendix~\ref{app:evaluation-details}.

\begin{figure}[htbp]
  \centering
  \includegraphics[width=\linewidth]{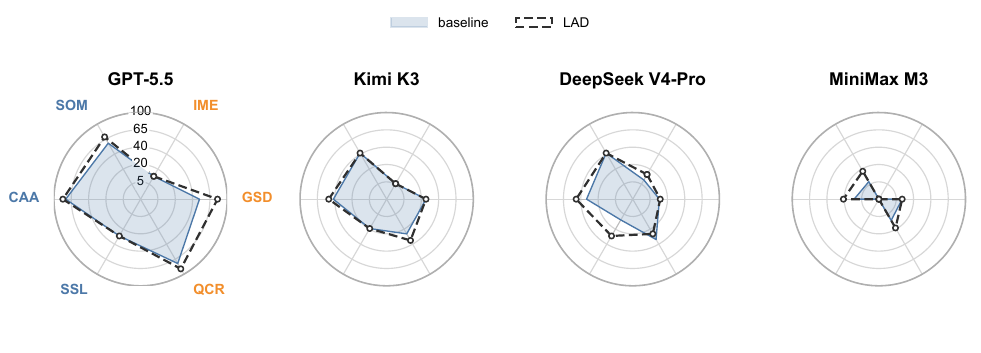}
  \caption{Difficulty-normalized field profiles for GPT-5.5, Kimi K3,
  DeepSeek V4-Pro, and MiniMax M3, from left to right.  The six axes represent
  the QAlg-Bench and QIT-Bench fields defined in Tables~\ref{tab:quantumalg-fields}
  and~\ref{tab:qit-fields}; exact values are reported in
  Table~\ref{tab:evaluation-fields}.  All panels use the same radial mapping,
  with equally spaced rings labelled 5, 20, 40, 65, and 100.  Baseline is shown
  as a translucent filled profile with a solid outline, while LAD is shown as
  an unfilled dashed profile with hollow nodes.
  Each axis aggregates 10 to 17 tasks, so a single verified task can
  visibly move a field score.}
  \label{fig:evaluation-fields}
\end{figure}

\subsection{Field-level capability profiles}
\label{sec:evaluation-fields}

We partition each suite into three fields and apply
Equation~\eqref{eq:objective-score} within each partition.  QAlg-Bench comprises
state and operator methods (SOM), circuit and algebraic algorithms (CAA), and
simulation, signal processing, and learning (SSL), with fixed difficulty
denominators 67, 105, and 93.  QIT-Bench comprises quantum channels and
representations (QCR), operator and state geometry, symmetry, and
distinguishability (GSD), and quantum information measures and entanglement
(IME), with denominators 80, 138, and 109.  These six disjoint partitions cover
the 76 tasks exactly.  Their unequal denominators mean that the same number of
additional verified tasks can produce different field-score changes, depending
on task difficulty and field size.
Each field contains between 10 and 17 tasks, and a single verified task
of difficulty 10 can move a field score by up to roughly 15 points at the
smallest denominator, so the field profiles should be read together with
these counts.

Figure~\ref{fig:evaluation-fields} shows that suite-wide scores mask persistent
field differences.  GPT-5.5 has the strongest observed profile overall, but
its QIT-Bench scores are much higher on QCR and GSD than on IME.  Across
models, SSL on QAlg-Bench and IME on QIT-Bench remain comparatively weak, so
observed verification success is
unevenly distributed across the fields in both suites.  The common radial
mapping permits axis-by-axis comparisons across panels, but polygon area is not
used in any aggregate or ranking.

Historical baseline--LAD differences are also localized.  GPT-5.5's largest
field change is a 30.4-score-point rise in GSD on QIT-Bench.  Kimi K3's CAA net
change is $+6.7$ score points, comprising two tasks verified only under LAD and
one task verified only under baseline.  Its QCR net change on QIT-Bench is $+8.8$ points
from one additional verified task and no regression; its other fields are
unchanged.  DeepSeek V4-Pro rises by 13.3 and 18.3 points in CAA and SSL on
QAlg-Bench but declines by 7.5 points in QCR on QIT-Bench.  MiniMax M3 remains
at 0.0 in SSL on QAlg-Bench and IME on QIT-Bench under both conditions.

The field profiles explain why a positive suite-wide change need not be
uniform.  DeepSeek V4-Pro's local QCR decline on QIT-Bench, for example,
coexists with a suite-wide QIT-Bench score change from 16.8 to 17.1 because
changes in other fields
offset it.  Figure~\ref{fig:evaluation-fields} should therefore be read as a
decomposition of Figure~\ref{fig:evaluation-summary} and
Table~\ref{tab:evaluation-paired}, with exact local values reported in
Appendix~\ref{app:evaluation-details}.
The following two sections describe the mathematical content behind these
field profiles and present representative tasks from each suite.
\section{Lean-QuantumAlg-Bench}
\label{sec:lean-quantumalg-bench}
To benchmark the proof capabilities of AI systems in quantum algorithms, we design QAlg-Bench, which contains 36 theorem-proving tasks drawn from the mathematics of quantum algorithms and uses the \textsc{Lean-QuantumAlg} library~\cite{zhang2026building,jing2026agentic}, developed by an overlapping subset of the present authors. The tasks are organized into three complementary fields, covering state and operator methods, circuit identities, number-theoretic algorithms, block encodings, and quantum signal processing. Each benchmark task specifies a Lean~4 theorem signature together with the supporting definitions available to the prover; success is determined by whether the submitted theorem body compiles in the fixed Lean environment.

\subsection{Task fields}
\label{sec:quantumalg-task-fields}

The benchmark groups eight mathematical scopes into three fields in Lean-QuantumAlg-Bench.
Table~\ref{tab:quantumalg-fields} summarizes their meanings and problem counts.

\begin{table}[htbp]
  \caption{    Task fields and scopes in QAlg-Bench.  }
  \label{tab:quantumalg-fields}
  \begin{tabular}{@{}lp{0.22\linewidth}p{0.48\linewidth}c@{}}
    \toprule
    \textbf{Field} & \textbf{Meaning} & \textbf{Scope} & \textbf{Count} \\
    \midrule
    SOM
      & state and operator methods
      & state, operator, measurement, and matrix algebra
      & 10 \\
    \cmidrule{1-4}
    CAA
      & circuit and algebraic algorithms
      & oracle and elementary quantum-circuit identities
      & 9 \\
    &
      & quantum Fourier transform, phase estimation, and order finding
      & 4 \\
    &
      & hidden subgroup and representation theory
      & 3 \\
    \cmidrule{1-4}
    SSL
      & simulation, signal processing, and learning
      & product formulas and norm bounds
      & 3 \\
    &
      & linear combinations of unitaries (LCU), block encoding, and quantum singular value transformation primitives
      & 2 \\
    &
      & quantum signal processing, parameter-shift rules, and trace estimation
      & 4 \\
    &
      & parameterized circuits, Lie algebra, and trainability
      & 1 \\
    \midrule
    & & \textbf{Total} & \textbf{36} \\
    \bottomrule
  \end{tabular}
\end{table}

The three fields emphasize complementary capabilities.
State and operator methods exercise matrix algebra, spectral arguments, and
reasoning about measurements.  Circuit and algebraic algorithms combine gate
composition, Fourier methods, order finding, and hidden-subgroup arguments.
Simulation, signal processing, and learning covers product-formula bounds,
block encodings, quantum signal processing, and quantum singular value
transformation~\cite{low2017optimal,gilyen2019quantum}, together with parameterized-circuit
trainability~\cite{mcclean2018barren}.  These fields exercise different proof skills,
including compositional tensor-product reasoning, number-theoretic
preconditions, and ancilla-register management.
The full task inventory appears in Appendix~\ref{app:full-benchmark-inventories}.

\subsection{Representative problems}
\label{sec:quantumalg-representative-problems}

We present three tasks chosen to illustrate distinct mathematical interfaces.
For each, we give the source-level statement, the Lean~4 formalization,
and the main issues a prover must navigate.

\subsubsection*{Problem 1: Walsh--Hadamard decoding of a Boolean character
  (circuit and algebraic algorithms)}

\paragraph{Source statement.}
The Bernstein--Vazirani algorithm~\cite{bernstein1997quantum} uses this
Walsh--Hadamard decoding identity.
For a fixed bitstring $s\in\{0,1\}^n$, define the Boolean character state
\begin{align*}
  \ket{\chi_s}
    &= 2^{-n/2}\sum_{y\in\{0,1\}^n}(-1)^{s\cdot y}\ket{y},
\end{align*}
where $s\cdot y=\sum_j s_j y_j \bmod 2$.  Prove that
\begin{align*}
  H^{\otimes n}\ket{\chi_s} &= \ket{s}.
\end{align*}

\paragraph{Formalization.}
The Lean statement (Listing~\ref{lst:walsh-hadamard}) is concise: one
bitstring argument and one equation.  Its mathematical structure is character
orthogonality over~$\mathbb{F}_2^n$.

\begin{leanlisting}{Walsh--Hadamard decoding of a Boolean character}{lst:walsh-hadamard}
namespace QAlgBench.WalshHadamardBooleanCharacter
noncomputable section
open QAlgBench Gate
variable {n : ℕ}

/-- H^{⊗n} |χ_s⟩ = |s⟩. -/
theorem main (s : Fin n → Bool) :
    HilbertOperator.applyVec hadamardTensorOp (chiState s)
      = (PureState.ket (R := BitReg n) s
           : StateVector (BitReg n)) := by
  sorry
\end{leanlisting}

The statement is short, but it already exercises library-specific skills.  The
prover must use the domain types \texttt{BitReg~n} (an $n$-qubit register),
\texttt{StateVector}, and \texttt{PureState.ket}, and handle the coercion from
the pure-state constructor to the ambient state-vector type.  Operator
application goes through \texttt{HilbertOperator.applyVec}, so the prover must
operate at the correct API layer above bare matrix multiplication.

\subsubsection*{Problem 2: Eigenvectors of modular multiplication in
  Shor's algorithm (circuit and algebraic algorithms)}

\paragraph{Source statement.}
Modular multiplication supplies the periodic unitary underlying Shor's
order-finding algorithm~\cite{shor1994algorithms}.  Kitaev's
eigenvalue-measurement formulation~\cite{kitaev1995quantum} supplies the
corresponding phase-estimation viewpoint.
Let $x,N\in\mathbb{N}^+$ with $x<N$ and $\gcd(x,N)=1$.  Let $r$ be
the multiplicative order of $x$ modulo~$N$.  Define
$U_x\ket{y}=\ket{xy\bmod N}$ and, for $s\in\{0,\dots,r-1\}$,
$$
  \ket{\psi_s}=\frac{1}{\sqrt{r}}\sum_{k=0}^{r-1}
    e^{-2\pi i sk/r}\ket{x^k\bmod N}.
$$
Prove that $\ket{\psi_s}$ is an eigenvector of $U_x$ with eigenvalue
$e^{2\pi i s/r}$.

\paragraph{Formalization.}
In Lean (Listing~\ref{lst:shor-eigenvectors}) the two-line textbook statement
expands to a theorem with six explicit hypotheses and three data arguments.

\begin{leanlisting}{Eigenvectors of modular multiplication in Shor's algorithm}{lst:shor-eigenvectors}
namespace QAlgBench.EigenvectorsModularMultiplicationShor
noncomputable section
open QAlgBench
variable {N : ℕ} [NeZero N]

theorem main (x : ℕ)
    (hx_pos : 0 < x) (hx_lt : x < N)
    (hcoprime : Nat.Coprime x N)
    (hunit : IsUnit (x : ZMod N))
    (r : ℕ) (hr : r = orderOf hunit.unit)
    (s : ℕ) (hs : s < r) :
    HilbertOperator.applyVec (UxOp (N := N) x)
      (psiS (N := N) x r s) =
      Complex.exp (2 * Real.pi * Complex.I
        * (s : ℂ) / (r : ℂ))
        • psiS (N := N) x r s := by
  sorry
\end{leanlisting}

The problem combines number-theoretic and quantum-mechanical reasoning.  The
prover must use Mathlib's \texttt{Nat.Coprime}, \texttt{ZMod~N},
\texttt{IsUnit}, and \texttt{orderOf} alongside the quantum types
\texttt{HilbertOperator} and \texttt{psiS}, and negotiate the coercion
$\mathbb{N}\to\texttt{ZMod~N}\to\texttt{Units}$.  The hypothesis \texttt{hunit}
(that $x$ is a unit in $\mathbb{Z}/N\mathbb{Z}$) is derivable from
coprimality but appears explicitly because \texttt{orderOf} requires a named
unit.  The remaining hypotheses (positivity, the bound $x<N$, coprimality, the
order definition, and the index bound) are implicit in the textbook phrase
``let~$r$ be the order of~$x$ mod~$N$.''

\subsubsection*{Problem 3: LCU block encoding from PREPARE and SELECT
  (simulation, signal processing, and learning)}

\paragraph{Source statement.}
This problem isolates an algebraic step in the method of linear combinations
of unitaries~\cite{childs2012hamiltonian}.  The circuit composed of PREPARE, SELECT,
and UNPREPARE appears in Hamiltonian simulation by a truncated Taylor
series~\cite{berry2015simulating}.  The resulting unitary is a
block encoding whose projected block equals $A/\alpha$~\cite{gilyen2019quantum}.
Let $\{\alpha_\ell>0\}$ be positive coefficients and $\{U_\ell\}$
unitaries on a system Hilbert space, and assume that the label states exhaust
the ancilla label basis.  Define
$A=\sum_\ell\alpha_\ell U_\ell$, $\alpha=\sum_\ell\alpha_\ell$, and
$\mathrm{SELECT}=\sum_\ell\proj{\ell}\otimes U_\ell$.  Let a PREPARE
unitary $P$ satisfy
$P\ket{0}=\alpha^{-1/2}\sum_\ell\sqrt{\alpha_\ell}\ket{\ell}$.
Set $W=(P^\dagger\otimes I)\,\mathrm{SELECT}\,(P\otimes I)$.
Prove that
$(\bra{0}\otimes I)\,W\,(\ket{0}\otimes I)=A/\alpha$.

\paragraph{Formalization.}
The Lean statement (Listing~\ref{lst:lcu-block-encoding}) manages an
ancilla--system tensor product, an embedding of the
LCU label set into the ancilla basis, and the projected-block extraction that
defines the block encoding.

\begin{leanlisting}{LCU block encoding from PREPARE and SELECT}{lst:lcu-block-encoding}
namespace QAlgBench.LCUBlockEncodingPrepareSelect
noncomputable section
open QAlgBench
open scoped BigOperators
variable {a n : ℕ} {L : Type*} [Fintype L] [DecidableEq L]

theorem ProjectedBlock.main
    (embed : L ↪ Fin (2 ^ a))
    (αcoeff : L → ℝ) (U : L → HilbertOperator (Qubits n))
    (P : Gate (Qubits a)) (W : Gate (Qubits (a + n)))
    (hαcoeff_pos : ∀ ℓ, 0 < αcoeff ℓ)
    (hU_unitary : ∀ ℓ, U ℓ ∈ Matrix.unitaryGroup (Fin (2^n)) ℂ)
    (hPrepare : ∀ j : Fin (2 ^ a),
      (P : HilbertOperator (Qubits a)) j 0 =
        ((Real.sqrt (coeffSum αcoeff))⁻¹ : ℝ) *
          ∑ ℓ, if embed ℓ = j
               then (Real.sqrt (αcoeff ℓ) : ℝ) else 0)
    (hW_eq : (W : HilbertOperator (Qubits (a + n))) =
      HilbertOperator.tensor
        (P : HilbertOperator (Qubits a)).conjTranspose
        (1 : HilbertOperator (Qubits n))
      * selectOp embed U
      * HilbertOperator.tensor
          (P : HilbertOperator (Qubits a))
          (1 : HilbertOperator (Qubits n))) :
    projectedBlock a n (W : HilbertOperator (Qubits (a + n))) =
      ((coeffSum αcoeff : ℂ)⁻¹) • weightedSum αcoeff U := by
  sorry
\end{leanlisting}

The benchmark-local \texttt{selectOp} vanishes outside the image of the label
embedding.  The theorem also assumes that $W$ is a unitary gate equal to the
PREPARE--SELECT composition.  These hypotheses are therefore jointly
realizable only when the embedded labels exhaust the ancilla basis.  The
formal statement proves the projected-block identity under this boundary;
it does not construct the usual unitary identity extension on unused labels.

The type \texttt{Qubits~(a~+~n)} encodes a composite Hilbert space of
dimension $2^{a+n}$, and its tensor-product structure is accessed through
\texttt{HilbertOperator.tensor}, abstracting away the explicit Kronecker
product.  The PREPARE hypothesis is stated as a column-zero condition on the
unitary matrix, so the prover must connect the gate-level and matrix-level
descriptions.  The label embedding $L\hookrightarrow\mathrm{Fin}(2^a)$ is
formalized with Lean's \texttt{Function.Embedding}, forcing the prover to
reason about injective maps between finite types.  The task consequently
combines quantum-algorithm content, linear algebra, and nontrivial use of
Lean's type system.

These three examples illustrate why QAlg-Bench does not reduce to one proof idiom.
Walsh--Hadamard decoding emphasizes state and operator interfaces, modular
multiplication combines quantum structure with number-theoretic coercions and
hypotheses, and LCU block encoding adds tensor composition and finite-type
embeddings.  Their different interfaces provide mathematical context for the
field-level results in Section~\ref{sec:evaluation}, where suite-wide results
are decomposed to show which mathematical and library interfaces remain
difficult.
\section{Lean-QIT-Bench}
\label{sec:lean-qit-bench}
To benchmark the proof capabilities of AI systems in quantum information theory, we design QIT-Bench, which contains 40 theorem-proving tasks drawn from quantum information theory and uses the \LeanQIT{} library~\cite{leanqit}, developed by an overlapping subset of the present authors. Its three fields cover channel representations; operator and state geometry, symmetry, and distinguishability; and information measures and entanglement. Each benchmark task specifies a Lean~4 theorem signature together with the supporting definitions available to the prover; success is determined by whether the submitted theorem body compiles in the fixed Lean environment.

\subsection{Task fields}
\label{sec:qit-task-fields}

The benchmark groups five mathematical scopes into three fields in Lean-QIT-Bench.
Table~\ref{tab:qit-fields} summarizes their meanings and problem counts.

\begin{table}[htbp]
  \caption{    Task fields and scopes in QIT-Bench.  }
  \label{tab:qit-fields}
  \begin{tabular}{@{}lp{0.22\linewidth}p{0.48\linewidth}c@{}}
    \toprule
    \textbf{Field} & \textbf{Meaning} & \textbf{Scope} & \textbf{Count} \\
    \midrule
    QCR
      & quantum channels and representations
      & channel, Choi, Kraus, and Stinespring-style representations
      & 11 \\
    \cmidrule{1-4}
    GSD
      & operator and state geometry, symmetry, and distinguishability
      & mixed-unitary channels, Werner--Holevo maps, and symmetry projectors
      & 7 \\
    &
      & trace norm, fidelity, continuity bounds, and gentle measurement
      & 10 \\
    \cmidrule{1-4}
    IME
      & quantum information measures and entanglement
      & entropy, source coding, Holevo/Fano bounds, and convexity
      & 9 \\
    &
      & entanglement conversion and hypothesis-testing tools
      & 3 \\
    \midrule
    & & \textbf{Total} & \textbf{40} \\
    \bottomrule
  \end{tabular}
\end{table}

The three fields emphasize complementary forms of reasoning.  Quantum
channels and representations asks the prover to move
among Kraus, Choi, Stinespring, and axiomatic \cptp{} descriptions.  Operator
and state geometry, symmetry, and distinguishability combines symmetry
computations with trace norm, fidelity, continuity bounds, and gentle
measurement.  Quantum information measures and entanglement covers entropy,
mutual information, and finite-resource information
measures~\cite{wilde2017quantum,tomamichel2016quantum}; the Holevo and Fano
bounds~\cite{wilde2017quantum}; and entanglement
concentration~\cite{bennett1996concentrating}.  The corresponding proof skills
include finite-index bookkeeping, combinatorial dimension formulas, chains of
operator inequalities, spectral majorization, and asymptotic
$\varepsilon$/$\delta$ reasoning over local operations and classical
communication (LOCC) protocols.  The full task inventory
appears in Appendix~\ref{app:full-benchmark-inventories}.

\subsection{Representative problems}
\label{sec:qit-representative-problems}

We present three tasks chosen to exercise distinct prover capabilities:
type-level index bookkeeping, real and operator inequality chaining, and
asymptotic resource-theoretic reasoning.  For each, we give the source-level
statement, the Lean~4 formalization, and the main issues a prover must
navigate.

\subsubsection*{Problem 1: Partial trace as a completely positive
  trace-preserving map (quantum channels and representations)}

The partial trace connects composite-system states to reduced states and
channel representations.  This task isolates the translation of an abstract
linear map into an explicit Kraus decomposition~\cite{wilde2017quantum}.  Its statement is quantified
over two arbitrary finite types \texttt{a} and \texttt{b}, so a valid proof
must hold uniformly across finite bipartite dimensions rather than for one
fixed matrix size.

\paragraph{Source statement.}
Let $\cH_A$ and $\cH_B$ be finite-dimensional Hilbert spaces.  The partial
trace over~$B$ is the linear map
$$
  \tr_B:\cL(\cH_A\otimes\cH_B)\to\cL(\cH_A).
$$
Prove that $\tr_B$ is completely positive and trace-preserving by giving an
explicit Kraus representation after fixing an orthonormal basis of~$\cH_B$.

\paragraph{Formalization.}
The Lean statement (Listing~\ref{lst:partial-trace-cptp}) packages the partial
trace as a \texttt{MatrixMap}, defines Kraus operators indexed by the
environment type~\texttt{b}, and asserts complete positivity through the Kraus
form together with trace preservation through the library predicate.

\begin{leanlisting}{Partial trace as a completely positive trace-preserving
  map}{lst:partial-trace-cptp}
namespace QITBench.PartialTraceCompletelyPositiveTracePreservingMap
noncomputable section
open QITBench

variable {a b : Type*} [Fintype a] [DecidableEq a] [Fintype b] [DecidableEq b]

noncomputable def partialTraceBMap : MatrixMap (a × b) a where
  toFun := partialTraceB (a := a) (b := b)
  map_add' X Y := partialTraceB_add X Y
  map_smul' c X := partialTraceB_smul c X

noncomputable def partialTraceBKraus (k : b) : Matrix a (a × b) ℂ :=
  fun i p => if p.1 = i ∧ p.2 = k then 1 else 0

/-- The partial trace over B is a CPTP map with explicit Kraus operators. -/
theorem main :
    partialTraceBMap (a := a) (b := b) =
        MatrixMap.ofKraus (partialTraceBKraus (a := a) (b := b)) ∧
      MatrixMap.IsTracePreserving (partialTraceBMap (a := a) (b := b)) := by
  sorry

end
end QITBench.PartialTraceCompletelyPositiveTracePreservingMap
\end{leanlisting}

The mathematics is elementary, and the real demand is organizational.  The
prover must reconcile three views of one object: the partial trace as a
concrete sum over an environment index, as a linear map between matrix spaces,
and as a Kraus decomposition.  Complete positivity is discharged through
\texttt{MatrixMap.ofKraus} applied to the explicit operators
\texttt{partialTraceBKraus}, and trace preservation through the separate
predicate \texttt{MatrixMap.IsTracePreserving}; the two views are then
reconciled by pointwise agreement on matrix entries.  The indexing
type~\texttt{b} stands in for the chosen orthonormal basis of~$\cH_B$, so the
proof is parameterized by an arbitrary finite type.  The problem therefore
tests index bookkeeping over product types and the discipline of staying at
the library's \texttt{MatrixMap} abstraction above bare matrix multiplication.

\subsubsection*{Problem 2: Fuchs--van de Graaf inequalities
  (operator and state geometry, symmetry, and distinguishability)}

The Fuchs--van de Graaf inequalities~\cite{fuchs1999cryptographic} relate trace distance and
root fidelity and
are standard tools in continuity and distinguishability arguments.  This task
probes a capability distinct from Problem~1: chaining real and operator
inequalities through functional calculus on complex matrices while working
directly with square roots and traces.  The statement is a conjunction, and
its two directions impose different proof obligations on the trace norm.

\paragraph{Source statement.}
For density operators $\rho,\sigma\in\cS(\cH)$, define the trace distance and
use the root-fidelity convention
$$
  D(\rho,\sigma)=\frac12\|\rho-\sigma\|_1,
  \qquad
  F(\rho,\sigma)=\tr\!\sqrt{\sqrt{\rho}\,\sigma\sqrt{\rho}}.
$$
Prove the two-sided bounds:
$$
  1-F(\rho,\sigma)\le D(\rho,\sigma)\le\sqrt{1-F(\rho,\sigma)^2}.
$$

\paragraph{Formalization.}
The Lean statement (Listing~\ref{lst:fuchs-van-de-graaf}) defines the trace
norm, trace distance, and root fidelity explicitly from matrix square
roots and traces, then states the two-sided inequality as a conjunction.

\begin{leanlisting}{Fuchs--van de Graaf inequalities}{lst:fuchs-van-de-graaf}
namespace QITBench.FuchsVanDeGraafInequalities
noncomputable section
open QITBench
open scoped ComplexOrder MatrixOrder

variable {n : ℕ}

noncomputable def matrixSqrt (A : CMatrix (Fin n)) : CMatrix (Fin n) :=
  CFC.sqrt A

noncomputable def traceNorm (A : CMatrix (Fin n)) : ℝ :=
  Complex.re (Matrix.trace (matrixSqrt (A.conjTranspose * A)))

noncomputable def traceDistance (rho sigma : CMatrix (Fin n)) : ℝ :=
  (1 / 2 : ℝ) * traceNorm (rho - sigma)

noncomputable def unsquaredFidelity (rho sigma : CMatrix (Fin n)) : ℝ :=
  Complex.re (Matrix.trace (matrixSqrt (matrixSqrt rho * sigma * matrixSqrt rho)))

/-- 1 - F(rho,sigma) <= D(rho,sigma) <= sqrt(1 - F(rho,sigma)^2). -/
theorem main (rho sigma : State (Fin n)) :
    let D := traceDistance rho.matrix sigma.matrix
    let F := unsquaredFidelity rho.matrix sigma.matrix
    1 - F ≤ D ∧ D ≤ Real.sqrt (1 - F ^ 2) := by
  sorry

end
end QITBench.FuchsVanDeGraafInequalities
\end{leanlisting}

The inequalities are mathematically standard, and the formal difficulty is
concentrated in the functional calculus.  The prover must build the trace
norm, trace distance, and root fidelity directly from \texttt{CFC.sqrt}
and \texttt{Matrix.trace}, taking real parts of complex traces throughout.
Both bounds demand careful handling of positivity under the square root and of
the singular values of $\rho-\sigma$.  Standard proofs derive the lower bound
from a measurement-based classical variational characterization and obtain the
upper bound from purification, Uhlmann's theorem, and monotonicity of trace
distance~\cite{wilde2017quantum}.  The problem therefore tests sustained manipulation
of real and operator inequalities inside an explicit formalization of the
matrix square root.

\subsubsection*{Problem 3: Achievability of entanglement concentration
  (quantum information measures and entanglement)}

Entanglement concentration establishes an achievable LOCC rate for converting
many copies of a bipartite pure state into maximally entangled
pairs~\cite{bennett1996concentrating}.  The task combines tensor powers, a
product-Kraus separable-operation abstraction motivated by the source-level
LOCC requirement, maximally entangled states of a prescribed rank, and
convergence of a fidelity sequence, all under a rate strictly below the Schmidt
entropy.  It therefore tests reasoning about an infinite family of protocols
together with Mathlib's topology interface.

\paragraph{Source statement.}
Let
$$
  \ket{\psi}_{AB}=\sum_{x\in\mathcal X}\sqrt{p(x)}\ket{x}_A\ket{x}_B
$$
be a bipartite pure state with Schmidt coefficients $\{p(x)\}_{x\in\mathcal
  X}$.  Its entanglement entropy is
$$
  E(\psi)=S(\psi_A)=-\sum_{x\in\mathcal X}p(x)\log_2 p(x).
$$
Let $\ket{\Phi_M}_{AB}=M^{-1/2}\sum_{j=1}^M\ket{j}_A\ket{j}_B$ be the standard
maximally entangled state of Schmidt rank~$M$.  Prove that for every rate
$0\le R<E(\psi)$ there exists a sequence of LOCC channels
$\{\Lambda_n\}_{n\ge1}$ and positive integers $M_n=\lfloor2^{nR}\rfloor$ such
that
$$
  \lim_{n\to\infty}
  F\!\left(
    \Lambda_n\!\left(\proj{\psi_{AB}}^{\otimes n}\right),
    \proj{\Phi_{M_n}}
  \right)=1,
$$
where $F(\rho,\sigma)=\|\sqrt{\rho}\sqrt{\sigma}\|_1$ denotes the quantum
fidelity.

\paragraph{Formalization.}
The Lean statement (Listing~\ref{lst:entanglement-concentration}) quantifies
over a probability distribution, a
bipartite pure state with prescribed Schmidt coefficients, a rate below the
Schmidt entropy, a target-rank function, and a family represented by
\texttt{SEPProtocol}.  This type stores a product-Kraus separable operation.
Every finite-round LOCC channel induces such a representation, but a general
separable operation need not admit a finite-round LOCC
implementation~\cite{wilde2017quantum,chitambar2014everything}.
Consequently, the formal target is a relaxation of the source achievability
claim: it verifies convergence within the broader separable-operation class.
The benchmark definition \texttt{schmidtEntropy} uses $\log_2$, and
\texttt{targetRankAtRate R n} is $\lfloor 2^{nR}\rfloor$, so the entropy,
rate, and target rank use the same bit convention.  The library function
\texttt{quantumFidelity} implements the same root-fidelity convention used in
Problem~2, rather than its square.

\begin{leanlisting}{Achievability of entanglement
  concentration}{lst:entanglement-concentration}
namespace QITBench.AchievabilityEntanglementConcentration
noncomputable section
open QITBench
open scoped BigOperators
open QITBench.OneShot

variable {X : Type*} [Fintype X] [DecidableEq X]

/-- For every rate R below the entanglement entropy, there is a sequence of
product-Kraus separable protocols whose output converges in fidelity to a
maximally entangled state of rank M_n = floor(2^{nR}). -/
theorem main
    (p : X → ℝ)
    (psi : PureVector (X × X))
    (hp : IsProbabilityDistribution p)
    (hpsi : HasSchmidtCoefficients psi p) :
    ∀ R : ℝ,
      0 ≤ R →
        R < schmidtEntropy p →
          ∃ M : ℕ → ℕ,
            ∃ protocols :
              (n : ℕ) →
                SEPProtocol
                  (TensorPower X n) (TensorPower X n)
                  (Fin (M n)) (Fin (M n)),
              (∀ n, 0 < M n) ∧
                (∀ n, M n = targetRankAtRate R n) ∧
                  Filter.Tendsto
                    (fun n =>
                      quantumFidelity
                        ((protocols n).channel.applyState
                          (psi.state.tensorPowerBipartite n)).matrix
                        (maximallyEntangledDensity (M n)))
                    Filter.atTop
                    (nhds 1) := by
  sorry

end
end QITBench.AchievabilityEntanglementConcentration
\end{leanlisting}

The prover must
assemble several abstractions from \LeanQIT{}: tensor powers of a bipartite
state through \texttt{psi.state.tensorPowerBipartite}, the product-Kraus/
separable-operation type \texttt{SEPProtocol}, the maximally entangled density
matrix \texttt{maximallyEntangled\allowbreak Density}, and the \texttt{quantumFidelity}
between each protocol output and its target.  The type
\texttt{SEPProtocol (TensorPower X n) (TensorPower X n) (Fin (M n)) (Fin (M n))}
fixes the input and output spaces of the $n$-th protocol, and
\texttt{targetRankAtRate R n} supplies $M_n=\lfloor2^{nR}\rfloor$ explicitly.
The conclusion is phrased with \texttt{Filter.Tendsto}, so the proof combines
quantum-information reasoning with Mathlib's topology library.  The statement
itself is an existential over an infinite family of protocols, and the proof
must reason about every family member uniformly.  The task illustrates the
suite's inclusion of resource-theoretic asymptotic statements alongside
finite-dimensional identities.

These examples likewise span distinct \LeanQIT{} interfaces, from finite-dimensional
channel representations to operator inequalities and asymptotic
resource-theoretic statements.  Their obligations combine indexed finite
types, matrix and operator norms, tensor powers, topology, and quantified
protocol families.  This diversity provides mathematical context for the
field-level scores reported alongside the suite-wide results in
Section~\ref{sec:evaluation}.
\section{Discussion and Conclusion}
\subsection{Discussions on the limitations}
\label{sec:limitations}
The reported results have four main validity limits that warrant further study: corpus coverage and difficulty calibration, semantic fidelity of the formal statements, comparability of the recorded runs, and availability of economic-cost records. Together, these limits constrain how far the observed scores and costs can be generalized beyond the present evaluation. These limitations make the reported results an initial measurement of the recorded benchmark runs, rather than a population estimate or a causal comparison of access conditions.

At the corpus level, the two suites sample theorem-sized tasks from selected parts of quantum algorithms and quantum information under the current Lean library interfaces. Their 36 and 40 tasks do not provide complete coverage of either domain, and the six author-defined fields aggregate areas of unequal size and maturity. Task selection, theorem granularity, and the availability of reusable definitions all affect what can enter the benchmark. Difficulty is assigned before model execution through two rubric-based reviews covering proof structure, theory dependencies, Lean engineering, and mathematical technique; scores differing by more than one point are adjudicated. This procedure prevents result-dependent reweighting, but the 1--10 assignments are not objective measurements of proof complexity. Some benchmark statements may overlap with material in model training corpora, and the present evaluation cannot determine such overlap. miniCTX~\cite{minictx} uses recency-based theorem selection to mitigate contamination; this benchmark does not.

Beyond corpus coverage, mechanical validity does not eliminate semantic risk. Refs.~\cite{ospanov2025minif2flean,ammanamanchi2026faults} show that kernel acceptance alone does not establish fidelity to the intended informal problem or robustness of the evaluation harness. In this benchmark, for example, the entanglement-concentration task represents protocols by product-Kraus separable operations rather than a full finite-round LOCC syntax, so its formal target is a relaxation of the motivating achievability claim. Lean compilation and automated checks apply to every task, while the separate semantic comparison described in Section~\ref{sec:quality-assurance-semantic-validation} covers only selected statements and does not support a benchmark-wide semantic-validity rate. Residual translation risk therefore remains.

For the fixed task set, the evaluation is bounded by the tested models, agent tools, budgets, hardware, suite versions, and access policies. Each suite--model--condition combination contains one run, so the results provide no estimate of stochastic variation. Historical baseline and LAD executions are not fully matched, which limits paired interpretation even under a common scoring rule. LAD adds a fixed reference library available only for consultation, but its semantic overlap with the benchmark targets is unmeasured. These results therefore do not characterize all theorem provers, retrieval systems, or forms of tool access.

Economic-cost accounting is also incomplete for some combinations because historical runs did not consistently retain the usage and pricing information needed to reconstruct API list-price-equivalent cost. Score, completion, and model invocation time remain available for every combination, but missing economic values limit cost comparisons. Provider-specific pricing assumptions and historical execution differences impose further constraints.

\subsection{Conclusion}
\label{sec:conclusion}
In this work, we introduced QAlg-Bench and QIT-Bench, two formal theorem-proving benchmarks for quantum algorithms and quantum information, comprising 76 tasks across six fields. A shared task contract, difficulty weights fixed before model execution, and Lean-based acceptance checks together define a rigorous and fully reproducible evaluation protocol that can be applied across different mathematical domains and library interfaces. Across four models and two access conditions, the highest observed scores are 60.4 on QAlg-Bench and 59.6 on QIT-Bench, showing that current models still struggle with certain structured quantum-theorem tasks. Verified proofs are unevenly distributed across fields, and the accompanying monetary and wall-clock costs provide a complementary view of model performance beyond pass rates alone. These single-run baselines establish an initial reference point for research on agentic provers in quantum information science and identify library access as a nontrivial variable whose effects warrant further study.

Our findings motivate matched repeated runs, broader task coverage, and more systematic semantic review. More importantly, the field-level weaknesses uncovered here make a strong case for proof agents that combine quantum-domain reasoning with more reliable use of library definitions and references. Taken together, QAlg-Bench and QIT-Bench provide a structured and extensible foundation for measuring and advancing the formal-proof capabilities of AI systems in quantum science. We expect these benchmarks to support the development of more capable and reliable proof agents. Moreover, they can guide the improvement of AI systems through fine-tuning, post-training, and harness-system design, paving the way toward self-evolving AI scientists for advancing quantum information science.

\section*{Acknowledgment}
This work was partially supported by the National Natural Science Foundation of China (Grant Nos.~92576114, 12447107) and the Guangdong Provincial Quantum Science Strategic Initiative (Grant Nos.~GDZX2403008, GDZX2503001, and GDZX2403001).

\clearpage

\appendix
\section{Benchmark settings}
\label{app:full-benchmark-inventories}

The following inventories list the 36 QAlg-Bench tasks and 40 QIT-Bench tasks used in
this study.  Each row records its field assignment, title, and difficulty.
Difficulty is a model-assisted integer assignment from 1 to 10 that was frozen
before model execution.  Two separate rubric reviews score proof structure,
theory dependencies, Lean engineering, and mathematical technique on
four axes scored from 1 to 4; the combined rubric maps these assessments to
the 1--10 scale, and a difference greater than one point receives adjudication.
The resulting value is used only as the weight $d_i$ in
Equation~\eqref{eq:objective-score}; it is not an intrinsic measure of proof
complexity and is not inferred from model performance.
The tables are generated deterministically from the fixed benchmark inventory,
whose difficulty values agree with the evaluation data used in this paper.

\begingroup
\begin{table}[tp]
\caption{QAlg-Bench task inventory.}
\label{tab:inventory-qalg}
\renewcommand{\arraystretch}{0.94}
\begin{tabular}{@{}llr@{}}
\toprule
Field & Title & Difficulty \\
\midrule
SOM & Spectral projectors of a Boolean phase oracle & 6 \\
 & Orthonormality of coherent Markov-chain rows & 5 \\
 & Unitary extension of a Markov-chain isometry & 9 \\
 & Post-measurement state after observing one qubit & 2 \\
 & Linear dependence over R versus over F2 & 6 \\
 & Fidelity of a Hamming-weight-truncated phase state & 10 \\
 & Weyl-average sanity check for the shift operator & 10 \\
 & Hilbert--Schmidt invariance under transposition & 4 \\
 & Trivial common unitary stabilizer of two full-rank states & 10 \\
 & Local marginal after an unread projective measurement & 5 \\
\cmidrule{1-3}
CAA & Unitarity and Hermiticity of the bit oracle & 2 \\
 & Matrix of H tensor H & 3 \\
 & Controlled phase from CNOT and Hadamards & 6 \\
 & Walsh--Hadamard decoding of a Boolean character & 9 \\
 & Controlled unitary invariant subspaces & 3 \\
 & Vanishing Pauli-Y expectation for real circuits & 7 \\
 & Exact QROM factorization into commuting address-selective operators & 9 \\
 & Phase kickback for a controlled eigenunitary & 4 \\
 & Two-CNOT synthesis of a controlled Ry rotation & 9 \\
 & Unitarity of the finite quantum Fourier transform & 8 \\
 & Product-state factorization of $\mathrm{QFT}_8\ket{3}$ & 6 \\
 & QPE with a superposition of exact eigenvectors & 7 \\
 & Eigenvectors of modular multiplication in Shor's algorithm & 10 \\
 & Weak Fourier sampling for a transposition in $S_n$ & 9 \\
 & Fidelity bound for hidden subgroup coset states & 3 \\
 & Rank of the strong Fourier sampling conditional state & 10 \\
\cmidrule{1-3}
SSL & Leading commutator term of a two-term exponential product & 9 \\
 & First-order Lie--Trotter global error scaling & 9 \\
 & Accumulation of product-formula errors for many small terms & 9 \\
 & Coefficient 1-norm in truncated Taylor simulation & 8 \\
 & LCU block encoding from PREPARE and SELECT & 9 \\
 & Postselected spectral transformation by trigonometric QSP & 10 \\
 & Shared-parameter shift rule for $R_X\otimes R_Z$ & 10 \\
 & Third moment from two controlled-SWAP tests & 9 \\
 & Parity and maximum of a Fejer-kernel QSP response & 10 \\
 & Approximate two-design depth and Haar loss fluctuations & 10 \\
\bottomrule
\end{tabular}
\end{table}
\endgroup
\begingroup
\begin{table}[tp]
\caption{QIT-Bench task inventory.}
\label{tab:inventory-qit}
\renewcommand{\arraystretch}{0.94}
\begin{tabular}{@{}llr@{}}
\toprule
Field & Title & Difficulty \\
\midrule
QCR & Unitarity of a System--Environment Evolution & 6 \\
 & Kraus Representation from a System--Environment Unitary & 7 \\
 & The Partial Trace as a Completely Positive Trace-Preserving Map & 6 \\
 & Trace Preservation in the Choi Representation & 5 \\
 & Unitality in the Choi Representation & 4 \\
 & Choi Matrix Acting on a Canonical Input State & 8 \\
 & Choi Positivity Characterization of Complete Positivity & 10 \\
 & A Fixed Point of Every Trace-Preserving Quantum Operation & 10 \\
 & Primitivity and Nondegeneracy of the Stationary Eigenvalue & 10 \\
 & A Partial-SWAP Marginal Splitter & 8 \\
 & Hadamard Channel from a Dephasing Isometry & 6 \\
\cmidrule{1-3}
GSD & The Choi Matrix of the Qutrit Werner--Holevo Channel & 7 \\
 & Trace Preservation and Unitality of the Werner--Holevo Channel & 4 \\
 & No Maximally Entangled Vector in the Qutrit Antisymmetric Subspace & 5 \\
 & The Symmetric Projector & 9 \\
 & The Antisymmetric Projector & 9 \\
 & Dimension of the Symmetric Subspace for Three Copies & 9 \\
 & Dimension of the Antisymmetric Subspace for Three Copies & 10 \\
 & Trace Norm and Unitary Optimization & 10 \\
 & Variational Characterization of the Trace Norm & 10 \\
 & Maximally Entangled State and the Hilbert--Schmidt Inner Product & 1 \\
 & Trace Distance Lower Bound for a Pure State & 10 \\
 & Trace-Distance Data Processing for Quantum Channels & 10 \\
 & Fuchs--van de Graaf Inequalities & 10 \\
 & Gentle Measurement Lemma for the Normalized Post-Measurement State & 10 \\
 & Alberti's Theorem in the Commuting Case & 10 \\
 & A Universal Upper Bound from Alberti's Theorem & 7 \\
 & Channel Fidelity and Gate Fidelity for a Pure Input State & 7 \\
\cmidrule{1-3}
IME & Negative Conditional Entropy and Entanglement for Pure Bipartite States & 10 \\
 & Random Unitary Realization of the Completely Depolarizing Channel & 10 \\
 & Uniqueness of the Maximum-Entropy State & 10 \\
 & Projective Measurement as a Random Unitary Channel & 6 \\
 & Entropy Increase under Non-Selective Projective Measurement & 8 \\
 & Convexity of Quantum Mutual Information & 9 \\
 & Holevo Bound and the Classical Capacity of $n$ Qubits & 10 \\
 & Quantum Fano Inequality & 10 \\
 & Spectrum and Entropy of a Single-Qubit Source & 8 \\
 & Exact Entanglement Dilution to a Maximally Entangled State & 8 \\
 & Achievability of Entanglement Concentration & 10 \\
 & Converse for Entanglement Concentration & 10 \\
\bottomrule
\end{tabular}
\end{table}
\endgroup

For each model in Table~\ref{tab:model-settings}, baseline and LAD use the same
model identifier, agent tool, and reasoning setting.  Each recorded task
receives one attempt with a 30-minute model invocation limit, and tasks
run serially in fresh isolated workspaces.  The agent may edit only the
theorem body and may run Lean on the current target, receiving the resulting
Lean output.  No partial conversation or answer is resumed, and external
network access is disabled.  If an infrastructure interruption requires
replacement, the replacement starts with a fresh model invocation.  After
every invocation, including a timeout, an independent Lean verification checks
the final submission.  The model and agent tool columns identify the evaluated
settings.  In Context window, an em dash marks runs without a separately pinned
long-context setting.

\begin{table}[t]
  \caption{Model and agent settings used for evaluation.}
  \label{tab:model-settings}
  \resizebox{\linewidth}{!}{
  \begin{tabular}{llllcc}
    \toprule
    Model & Developer & Model identifier & Agent tool &
      Reasoning setting & Context window \\
    \midrule
    GPT-5.5 & OpenAI & \nolinkurl{gpt-5.5} & Codex & xhigh & 256k \\
    Kimi K3 & Moonshot AI & \nolinkurl{k3[1m]} & Claude Code & max & 1M \\
    DeepSeek V4-Pro & DeepSeek & \nolinkurl{deepseek-v4-pro[1m]} & Claude Code & max & 1M \\
    MiniMax M3 & MiniMax & MiniMax M3 & Claude Code & max & \textemdash{} \\
    \bottomrule
  \end{tabular}
  }
\end{table}
\section{More experimental details}
\label{app:evaluation-details}

This appendix provides the absolute resource metrics, within-suite rankings,
and field values underlying the main evaluation.

Table~\ref{tab:evaluation-absolute} reports completion and the per-score cost
values behind Table~\ref{tab:evaluation-paired}.

\begin{table}[htbp]
\caption{Absolute completion and per-score cost metrics. Completion is the verified-problem percentage, and economic and time cost are reported in USD and seconds per score point, respectively. Economic cost divides mean API list-price-equivalent USD among tasks with reconstructable prices by the corresponding difficulty-weighted score. Time cost is the full-suite mean model invocation seconds per task divided by the overall score; model invocation time includes the complete agent invocation and Lean checks initiated by the agent but excludes the independent final verification. A black em dash denotes an unavailable observed metric.}
\label{tab:evaluation-absolute}
\begin{tabular}{@{}lllrrr@{}}
\toprule
Suite & Condition & Model & Completion & USD/score point & s/score point \\
\midrule
QAlg-Bench & baseline & GPT-5.5 & 61.1\% & 0.040530 & 9.607 \\
 &  & Kimi K3 & 47.2\% & 0.019408 & 34.806 \\
 &  & DeepSeek V4-Pro & 41.7\% & 0.001098 & 48.629 \\
 &  & MiniMax M3 & 8.3\% & \textemdash{} & 118.860 \\
\cmidrule{2-6}
 & LAD & GPT-5.5 & 69.4\% & 0.033640 & 7.729 \\
 &  & Kimi K3 & 50.0\% & 0.023276 & 32.678 \\
 &  & DeepSeek V4-Pro & 52.8\% & 0.000977 & 32.528 \\
 &  & MiniMax M3 & 25.0\% & \textemdash{} & 89.355 \\
\midrule
QIT-Bench & baseline & GPT-5.5 & 52.5\% & 0.049337 & 19.287 \\
 &  & Kimi K3 & 27.5\% & 0.046758 & 74.849 \\
 &  & DeepSeek V4-Pro & 22.5\% & 0.004859 & 88.594 \\
 &  & MiniMax M3 & 12.5\% & \textemdash{} & 172.318 \\
\cmidrule{2-6}
 & LAD & GPT-5.5 & 65.0\% & 0.042586 & 12.949 \\
 &  & Kimi K3 & 30.0\% & 0.045455 & 70.350 \\
 &  & DeepSeek V4-Pro & 25.0\% & 0.002895 & 85.048 \\
 &  & MiniMax M3 & 15.0\% & \textemdash{} & 135.409 \\
\bottomrule
\end{tabular}
\end{table}

\begin{figure}[htbp]
  \centering
  \includegraphics[width=\linewidth]{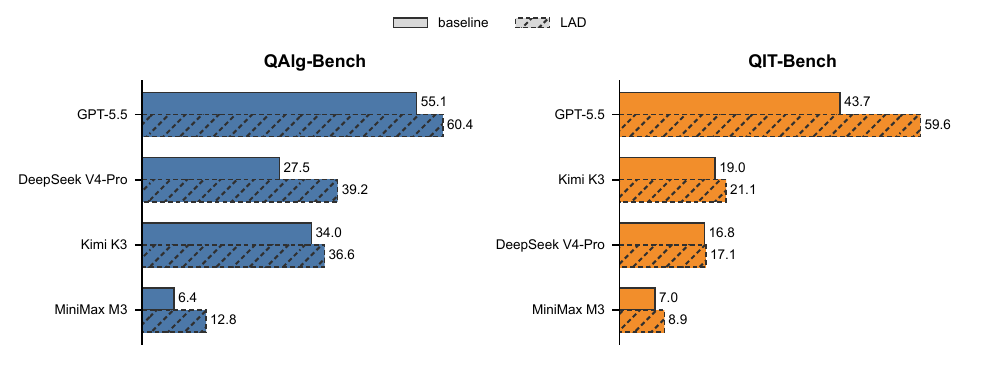}
  \caption{Within-suite score leaderboards for QAlg-Bench and QIT-Bench.  Each model has
  adjacent baseline and LAD bars, with exact difficulty-weighted scores printed
  at the bar ends.  Models are ordered within each suite by the larger of their
  two condition scores; ties are ordered as GPT-5.5, Kimi K3, DeepSeek V4-Pro,
  and MiniMax M3.  Blue and
  orange identify QAlg-Bench and QIT-Bench, while solid outlines identify baseline and
  dashed, diagonally hatched outlines identify LAD.}
  \label{fig:evaluation-leaderboard}
\end{figure}

Figure~\ref{fig:evaluation-leaderboard} presents the absolute scores as a
compact within-suite ranking.  Each model has adjacent baseline and LAD bars,
and the printed values provide the exact comparison.

Table~\ref{tab:evaluation-fields} gives the exact field values behind
Figure~\ref{fig:evaluation-fields}.  The QAlg-Bench and QIT-Bench subtables spell out
baseline and LAD for each model.  Their fixed difficulty denominators are 67,
105, and 93 for SOM, CAA, and SSL, and 80, 138, and 109 for QCR, GSD, and IME.
Within each suite, the three field partitions are disjoint and cover the
complete task set.

\begin{table}[htbp]
\caption{Difficulty-normalized field scores by suite and condition. Subtables (a,b) report QAlg-Bench and QIT-Bench, respectively. The fixed difficulty denominators are SOM=67, CAA=105, SSL=93 for QAlg-Bench and QCR=80, GSD=138, IME=109 for QIT-Bench.}
\label{tab:evaluation-fields}
\begin{subtable}[t]{0.49\linewidth}
\caption{QAlg-Bench}
\label{tab:evaluation-fields-qalg}
\resizebox{\linewidth}{!}{
\begin{tabular}{@{}llrrr@{}}
\toprule
Condition & Model & SOM & CAA & SSL \\
\midrule
baseline & GPT-5.5 & 58.2 & 76.2 & 29.0 \\
 & Kimi K3 & 41.8 & 41.9 & 19.4 \\
 & DeepSeek V4-Pro & 41.8 & 33.3 & 10.8 \\
 & MiniMax M3 & 7.5 & 11.4 & 0.0 \\
\cmidrule{1-5}
LAD & GPT-5.5 & 70.1 & 81.9 & 29.0 \\
 & Kimi K3 & 41.8 & 48.6 & 19.4 \\
 & DeepSeek V4-Pro & 41.8 & 46.7 & 29.0 \\
 & MiniMax M3 & 17.9 & 21.0 & 0.0 \\
\bottomrule
\end{tabular}
}
\end{subtable}
\hfill
\begin{subtable}[t]{0.49\linewidth}
\caption{QIT-Bench}
\label{tab:evaluation-fields-qit}
\resizebox{\linewidth}{!}{
\begin{tabular}{@{}llrrr@{}}
\toprule
Condition & Model & QCR & GSD & IME \\
\midrule
baseline & GPT-5.5 & 75.0 & 50.0 & 12.8 \\
 & Kimi K3 & 26.2 & 25.4 & 5.5 \\
 & DeepSeek V4-Pro & 33.8 & 13.0 & 9.2 \\
 & MiniMax M3 & 11.2 & 10.1 & 0.0 \\
\cmidrule{1-5}
LAD & GPT-5.5 & 87.5 & 80.4 & 12.8 \\
 & Kimi K3 & 35.0 & 25.4 & 5.5 \\
 & DeepSeek V4-Pro & 26.2 & 13.8 & 14.7 \\
 & MiniMax M3 & 18.8 & 10.1 & 0.0 \\
\bottomrule
\end{tabular}
}
\end{subtable}
\end{table}

Together, these values provide a numerical basis for checking the overall and
field-level comparisons in Section~\ref{sec:evaluation}.

\end{document}